\documentstyle[12pt]{article}
\begin{document}
\vspace{.5cm}
\begin{center}
{\Huge A Liquid Model Analogue\\ for\\ Black Hole Thermodynamics}
\end{center}
\vspace{.5cm}
\begin{center}
{ David Hochberg\footnote{Electronic address:
hochberg@laeff.esa.es} and Juan P\'erez-Mercader\footnote{Also at: IMAFF, CSIC, Madrid.}$^{,}$ \footnote{Electronic address:
mercader@laeff.esa.es}}\\
\vspace{0.5cm}
{\sl Laboratorio de Astrof\'{\i}sica Espacial y F\'{\i}sica 
Fundamental\\
Apartado 50727\\
28080  Madrid, Spain}
\end{center}
\vspace{-9cm}
\begin{flushright}
LAEFF-96/21\\
\end{flushright}
\vspace{7cm}
\begin{center}
\today
\end{center}
\begin{abstract} 
We are able to characterize a 2--dimensional classical 
fluid sharing some of the same
thermodynamic state functions as the Schwarzschild black hole. 
This phenomenological 
correspondence between black holes and fluids is established by means of the
model liquid's pair-correlation function and the two-body 
atomic interaction potential.
These latter two functions are calculated exactly in terms of the black hole
internal (quasilocal) energy and the isothermal compressibility. We find the existence of a ``screening" like effect for the components of the liquid.
\end{abstract}

\vfill\eject
\section{Introduction.}

The derivation of the thermal properties of a black hole is typically
carried out in the context of the 
Euclidean path integral approach, as initiated by 
Gibbons and Hawking \cite{GibHawk77}. In this language, which is 
manifestly geometrical,
the black hole partition function is identified with the Euclidean path
integral, the integration over four-metrics playing the role of the
configurational sum, and the hamiltonian, a function of the metric and 
curvature, taken to be that of the gravitational system.  

A black hole of mass $M$
has a temperature $T_{\infty} ={\hbar}/{(8 \pi M)}$ measured at large
distances from the hole. Following and extending this approach, York
demonstrated that the canonical ensemble with elements of radius $r$ and
temperature $T(r)$ for hot gravity with black holes is well-defined
\cite{York86}.
That is, one treats a collection of spherical cavities of radius $r$
with a temperature $T$ at $r$. These cavities may contain either, no black
hole, or one of two physically distinct black holes
depending on the value of the product
$rT$. In the case when the two distinct solutions pertains, only one of 
them will correspond to a thermodynamically stable black hole.  This
ensemble resolves a number of difficulties in assessing the physical
significance of the classical black hole action in its 
contribution to the Euclidean
path integral. 

However, the reasons for considering the implantation
of a black hole in a finite cavity, or ``box'', go well beyond the resolution
of these initial difficulties and in fact the spatial cut-off provided by the
cavity has been recognized as being crucial for making sense of black hole
thermodynamics in general, quite independent of the path integral
approach. For example, when one comes to consider the back-reaction of the
radiation on the spacetime geometry of the black hole, the system comprised
of black hole plus radiation must be put into a finite cavity, lest the
radiation in a spatial region that is too large collapses onto the hole,
thereby producing a larger one \cite{York85}.  
Related to this (but much more
general) is the fact that the usual thermodynamic limit in which one invokes
the infinite system volume limit does not exist for equilibrium,
self-gravitating systems
at a finite temperature. This follows since the system is unstable to 
gravitational collapse, or recollapse, if a black hole is already present.
This, in practice, presents no problem since
physically, one only requires that the system 
can in {\em principle} become
sufficiently large so that fluctuations become negligible. This peculiarity
of gravitational thermodynamics will play an important role in the present
paper.

While the Euclidean path integral approach is well-defined and allows one
to obtain the same value of the entropy as required by the thermodynamic
analogy, the Bekenstein-Hawking entropy, it does not shed any light on the
so-called dynamic origin of the entropy, nor does it explain the
``microphysics'' giving rise
to the macroscopic thermodynamic functions, such
as the internal energy, heat capacity, the equation-of-state, etc., that
characterize the black hole. This state of affairs has spawned numerous
efforts to understand the dynamical, or statistical mechanical, origins of
black hole thermodynamics with particular emphasis paid to an explanation
of the dynamical origin of 
black hole entropy as for example in Refs. \cite{York83} and
\cite{Zurek}-\cite{Barvinsky}.

In contrast to on-going efforts devoted to identifying the black hole's 
fundamental degrees of
freedom, we wish to take a model-oriented approach and promote 
a phenomenological $analogy$
between black holes and liquids. The analogy will be established at the level
of thermodynamics. 
In the present paper, we seek what might
be termed an effective {\em atomic} picture of black hole
thermodynamics. By this we mean that it may be possible to reproduce (some)
black hole thermodynamics in terms of microscopic properties of  
an interacting fluid or
gas. The components of this analogue fluid are massive point particles
interacting mutually via a pairwise additive potential. If such a
correspondence is possible, we will have effected a mapping between the
inherently geometrical degrees of freedom part and parcel of the Euclidean
approach, and the ``atomic'' variables actually
appearing in standard partition
functions for fluids.  The geometric quantities include
metric,
curvature, manifolds and boundaries. The so-called atomic 
quantities include the
particles, their momenta and positions, and their interaction
potentials. The correspondence is established via the black hole's
thermodynamic functions, as derived from the standard Euclidean 
path integral approach, using these as given input. The task is then to
characterize a liquid or (dense) gas whose microscopic properties (as encoded 
for example by the
potential, pressure, pair-correlation function)
can be ``tuned'', or adjusted suitably, so as
to reproduce mathematically the same set of black hole state functions.
If the program so described is successful, then one has in effect, replaced
integration over metrics by an integration over a multi-particle classical
phase space while the gravitational action is replaced by a particle
hamiltonian, containing a (non-relativistic) kinetic energy and potential
energy term.

In the next Section we review and comment on the essential features of 
black hole thermodynamics as derived from the Euclidean path integral in
the saddle point approximation. The black hole energy, entropy, equation of 
state and compressibility are displayed and their qualitative features are
revealed through various limits and graphical representations. The way in which
we establish a connection between liquids and black holes is taken up in
Section III. The key in building a ``liquid'' model 
is provided by the fundamental
equations employed in the study of the atomic dynamics 
of simple liquids and the
atomic picture of the thermodynamics of the liquid state. These fundamental
relations
equate the macroscopic (thermodynamic) to the microscopic (internal structure,
potential energy) properties of fluids. These relations are derived as rigorous
consequences of statistical mechanics applied to fluids and (dense) gases.
A particular type of fluid is singled
out the moment we identify the macroscopic variables of the fluid 
with those of the
black hole. The points of contact between the black hole and fluid are set up
via their respective internal energies and compressibilities. The analog fluid is 
identified to the extent that we can write down its pair-potential and 
two-body correlation function.
The, necessarily, bounded spatial extent of the black hole ensemble is crucial
in allowing us to solve for the liquid's microscopic parameters exactly.
These are calculated in closed form as well as presented graphically.

The ultimate purpose of establishing such a mapping is the double benefit
to be gained in being able to relate black hole 
physics to the molecular dynamics of
fluids.  Recent work of a similar spirit includes the possible
correspondence between black holes and quantized vortices 
in superfluids \cite{Volovik}
and a connection between fluid surface tension and black hole 
entropy \cite{Callaway}.
A summary is given in Section III. Absolute units $(G=c=\hbar=k_B=1)$ are used
throughout except where restoration of conventional units may be helpful.

\section{Black hole thermodynamics in brief.}

 In deriving gravitational thermodynamics from an Euclidean path integral
\begin{equation}
Z(\beta) = \int d{\mu}[g,\phi]\, e^{-I[g,\phi]} = e^{-\beta F},
\end{equation}
one expects the dominant contribution to the canonical partition function to 
come from those metrics $g$ and matter
fields $\phi$ that are near a background metric
$g^{(0)}$ and background field $\phi^{(0)}$, respectively. These background fields are obtained from 
solutions of the classical field equations. The classical contribution
to $Z$ is obtained by evaluating $Z$ at 
the point $(g^{(0)}, \phi^{(0)})$ in which
case one obtains the familiar relation
\begin{equation}
\beta F = I[g^{(0)}, \phi^{(0)}] = I[g^{(0)}],
\end{equation}
where we have taken $\phi^{(0)} = 0$ in the last equality. This provides
the free energy $F$ of the gravitational system 
in the saddle-point approximation.  
The action $I$ is the first-order Euclidean Einstein action including a
subtraction term necessary to avoid runaway solutions.
The action
appropriate to a black hole in a spherical cavity of radius $r$ is given by
\cite{York86}
\begin{equation}
I = I_1 - I_{subtract},
\end{equation}
where
\begin{equation}
I_1 = -\frac{1}{16 \pi}\int^{r}_{2M} \int^{\beta_*}_0 d^4x \, \sqrt{g}\, R
 + \frac{1}{8 \pi} \oint_{S^1 \times S^2} d^3x \, \sqrt{\gamma}\, Tr({\cal K})
\end{equation}
The action receives in general both volume and boundary contributions.
The Euclidean four-space metric is
\begin{equation}
g_{\mu \nu} = {\rm diag} \left((1-\frac{2M}{r}),(1-\frac{2M}{r})^{-1},
r^2, r^2 \sin^2 \theta \right).
\end{equation}
For this metric, the volume contribution vanishes identically.
The boundary at $r = const.$ is the product of $S^1 \times S^2$ of the 
periodically identified Euclidean time with the two-sphere of area
$A = 4\pi r^2$. The period of Euclidean time, identified with the 
$S^1$-coordinate, is
$\beta_* = 8\pi M$. The trace of the boundary extrinsic curvature is
denoted by $Tr({\cal K})$ and $\gamma$ is the induced 3-metric on the
boundary. Finally, $I_{subtract}$ is $I_1$ evaluated on a flat spacetime
having the same boundary $S^1 \times S^2$.

It is important to remember that for the canonical ensemble
the mass parameter $M$ appearing in
these formulae is not simply a
constant but is instead a specific function of
the cavity radius $r \geq 0$ and the cavity wall 
temperature $T(r)\geq 0$ \cite{York85,York86}. This can
be verified by inverting the expression for the blue-shifted temperature in
equation
(11) below and solving for $M = M(r,T)$. The relation so obtained is a 
cubic equation in $M$.
When $rT < \sqrt{27}/{8 \pi}=(rT)_{min} \approx 0.207 $, there are no real 
solutions of this equation.
On the other hand, when  $rT \geq
(rT)_{min} $, there exist two real and non--negative
branches given by
\begin{eqnarray}
M_2(r,T) &=& \frac{r}{6} \left[1 + 2 \cos (\frac{\alpha}{3}) \right],\\
M_1(r,T) &=& \frac{r}{6} \left[1 - 2 \cos (\frac{\alpha + \pi}{3}) \right],\\
\cos (\alpha) & = & 1 - \frac{27}{32 \pi^2 r^2 T^2}, \\
0 & \leq & \alpha \leq \pi.
\end{eqnarray}
This shows that the Schwarzschild mass is in fact
double-valued in the canonical ensemble. 
One has that $M_2 \geq M_1$, with equality holding at $rT = (rT)_{min}$.
The heavier mass branch, $M_2$, is the thermodynamically 
stable solution because it leads to the lowest free energy, Eq. (2), and is the
one we shall be considering in the remainder of this work.

Calculating the action $I$ from (3) and (4) yields $(I_{subtract} = -\beta r)$
\begin{equation}
I = 12 \pi M^2 - 8 \pi M r + \beta r,
\end{equation}
where 
\begin{equation}
\beta = T^{-1}(r) = 8\pi M \left( 1 - \frac{2M}{r} \right)^{1/2},
\end{equation}
is the inverse local temperature and is the proper length of the $S^1$ 
component of the boundary.
Employing $I$ and the saddle-point approximation $\beta F = I$, it is a 
straightforward exercise to calculate the thermodynamic state functions
associated with a black hole in the canonical ensemble. In so doing, it is 
useful to note the following two identities
\begin{eqnarray}
\left( \frac{\partial M}{\partial r} \right)_{\beta} & = & -
\frac{ \frac{M^2}{r^2} } { (1 - \frac{3M}{r}) }, \\
\left( \frac{\partial M}{\partial \beta} \right)_{A} & = & \frac{1}{8 \pi}
\frac{ (1 - \frac{2M}{r})^{1/2} }{ (1 - \frac{3M}{r}) },
\end{eqnarray}
which may be deduced from the expression for the inverse 
local temperature (11).
The black hole's internal, or thermal, energy is
\begin{equation}
 E = -\left( \frac{\partial \ln Z}{\partial \beta} \right)_{A} =
  \left( \frac{\partial I}{\partial \beta} \right)_{A} = r - 
 r \left(1 - \frac{2M}{r} \right)^{1/2}.
\end{equation}
The entropy $S$ is 
\begin{equation}
S = \beta \left( \frac{\partial I}{\partial \beta} \right)_{A} - I =
4\pi M^2 , 
\end{equation}
while the surface pressure $\sigma$ is
\begin{equation}
\sigma = -\left( \frac{\partial F}{\partial A} \right)_{T} =
\frac{1}{8\pi r} \left[ \frac{\left(1 - \frac{M}{r}\right)}
{\left(1 - \frac{2M}{r}\right)^{1/2}}
- 1 \right].
\end{equation}

Another quantity of special interest in what is to follow, is the black hole
isothermal compressibility, $\kappa_T (A)$, which again can be calculated
using the standard prescription,
\begin{eqnarray}
\kappa_T (A)& = & -\frac{1}{A} \left( \frac{\partial A}{\partial \sigma} 
\right)_T  =  16\pi r \left(\frac{r}{M}\right)^3 
\left(1 - \frac{3M}{r} \right)\left(1-\frac{2M}{r}\right)^{3/2}
/   \nonumber \\
\Bigl\{ 1 & + &  \left( \frac{r}{M} \right)^3 \left( \frac{3M}{r}-1 \right)
[ \left(1-\frac{2M}{r}\right)^{3/2} - 1 
 + \frac{3M}{r} - \frac{3M^2}{r^2} ] \Bigr\}.
\end{eqnarray}
Although at face 
value these functions appear to have a complicated dependence on $r$ and $T$,
they are actually quite simple, owing to their dependence on the 
slowly varying  ratio $M/r$. To gain some insight into the 
behavior of these functions, it
is useful to examine some limiting cases, namely when $ (i)\, rT \rightarrow 
\infty$ and when $(ii)\, rT \rightarrow (rT)_{min}$. The limit $(i)$ is understood to mean either that $r \rightarrow \infty$ or $T \rightarrow \infty$, or
both, simultaneously. The second limit $(ii)$ actually 
defines a hyperbola in the
$r-T$ plane along which the two independent limits 
$r \rightarrow 0 \,(T \rightarrow \infty)$
or $r \rightarrow \infty \, (T \rightarrow 0)$ can be taken.
The mass function takes the
form $M(r,T) \approx \frac{r}{2}(1 - \frac{1}{4\pi^2 r^2T^2})$, and 
$M(r,T) = \frac{r}{3}$, respectively. Physically, these limits indicate that
for all allowed values of $r$ and $T$, the cavity wall radius always lies 
between the black hole's event horizon $(r = 2M)$ and the unstable circular
photon orbit $(r = 3M)$. 
The behavior of the black hole
internal energy with respect to these limits is 
\begin{equation}
(i) \qquad E \rightarrow r - \frac{1}{(4\pi T)}, 
\end{equation}
and 
\begin{equation}
(ii) \qquad E = \frac{\sqrt{3} -1}{\sqrt{3}} r.
\end{equation}
$E$ is essentially a positive linear function of $r$, depending only
very weakly on the temperature for large values of $rT$. For $rT = (rT)_{min}$,
$E$ is strictly linear in $r$, or inversely proportional to $T$, 
since of course, in
this latter limit, $r \sim 1/T$. Note that on this hyperbola, 
$E \rightarrow 0$ for
$r \rightarrow 0$ (or $T \rightarrow \infty$). The equation for the
surface pressure is also an equation of state for the black hole
pressure since it is expressed as a function of 
the cavity radius $r$, which gives a measure of the system
size,  and the boundary temperature $T$: $\sigma = \sigma(r,T)$.
Using the limiting forms of the mass function, one can show that
the asymptotic limit of the surface pressure is 
\begin{equation}
(i) \qquad \sigma \rightarrow  \frac{T}{4} - \frac{1}{8\pi r},
\end{equation}
for $rT \rightarrow \infty$, so that the pressure increases with the
temperature, depending only very weakly on the system size. 
When evaluated along the limit hyperbola, one obtains
\begin{equation}
(ii) \qquad \sigma = \left( \frac{2\sqrt{3}}{3} -1 \right)
\frac{1}{8\pi r};
\end{equation}
in other words, in this regime, the pressure increases as the cavity 
radius (or area) decreases. Because of the reciprocal relation between $r$ and
$T$ along the hyperbola, this is equivalent to increasing pressure with
increasing temperature. Such qualitative behavior is familiar from the 
ideal gas. Finally, the limiting forms of the isothermal compressibility
are 
\begin{equation}
(i) \qquad \kappa_T(A) \rightarrow -16\pi r < 0,
\end{equation}
and 
\begin{equation}
(ii) \qquad \kappa_T(A) = 0.
\end{equation}
These two latter limits deserve some special comment. First, note that
the black hole isothermal compressibility is generally negative. This
is an unfamiliar property in regards to
 conventional thermodynamic systems. Indeed,
standard textbook arguments prove that $\kappa_T \geq 0$, irrespective of
the nature of the substance comprising the system. However, a key step
in those proofs assumes that quantities such as the temperature and pressure
are {\em intensive}, that is, independent of the size of the system and
constant throughout its interior. Such is most emphatically 
$not$ the case for
gravitating systems such as black holes, where in fact, the temperature
and pressure are not intensive quantities 
but are instead scale dependent. 
An equilibrium self-gravitating
object does not have a spatially constant temperature. This is a 
consequence of the principle of equivalence which implies that temperature 
is red- or blue-shifted in the same manner as the frecuency of photons in a 
gravitational field.  Secondly, for
values $rT = (rT)_{min}$, the compressibility vanishes identically.
This qualitative behavior is familiar from the classical picture of a solid
at $T=0$ (no density fluctuations $\Longrightarrow$ zero compressibility).

In the Figures 1-3, we have plotted $E,\sigma$ and $\kappa_T$ in the 
$r-T$ plane subject to the condition $rT \geq (rT)_{min}$. As indicated in 
Fig. 1, the black hole energy is a positive increasing function in $r$ and is
fairly insensitive to changes in the temperature for values of $T \geq 0.4$.
The flat region at zero level corresponds to the locus of excluded 
points satisfying
$rT < (rT)_{min}$ and is therefore not to be considered as part of the graph
as such. 
The cavity wall surface pressure, shown in Fig. 2, is a 
positive increasing function of $T$ and varies slowly in $r$ for $r \geq 0.5$.
The flat null region represents the same locus of points 
as in the prior graph. Finally,
the black hole isothermal compressibility is a 
{\em negative definite function} for
all $rT > (rT)_{min}$, decreasing for increasing $r$ and relatively constant
with respect to changes in the temperature.

Other functions that may be calculated include the specific heats at
constant area and at constant pressure, respectively, as well as the
adiabatic compressibility, but these are of no direct interest for the
present consideration.

Finally, to complete this brief overview of black hole thermodynamics, we
need to identify the effective spatial dimension of the system. It will not
have escaped the reader's attention that the above functions have been
defined and calculated in terms of the cavity wall area $A$, rather than in
terms of the cavity volume. Spatial volume is not well defined in the
presence of a black hole, whereas the area is, and it is the latter which
provides the correct means for measuring the size of the system \cite{York86}.
That the wall area is the proper extensive variable to use is confirmed by
considering the black hole's thermodynamic identity. The explicit
calculation of the following partial derivatives
\begin{equation}
\left( \frac{\partial E}{\partial S} \right)_A = \frac{1}{8\pi M}
\left( \frac{\partial E}{\partial M} \right)_A = \frac{1}{8\pi M}\left(1 - 
\frac{2M}{r}\right)^{-1/2} \equiv T,
\end{equation}
and
\begin{equation}
\left( \frac{\partial E}{\partial A} \right)_S = \frac{1}{8\pi r}
\left( \frac{\partial E}{\partial r} \right)_M = \frac{1}{8\pi r}
\left[ 1 -\frac{\left( 1 - \frac{M}{r} \right)}
 { \left(1 -\frac{2M}{r} \right)^{1/2} }
\right] \equiv -\sigma.
\end{equation}
proves that
\begin{equation}
dE = T\, dS - \sigma \, dA
\end{equation}
is an exact differential. In other words,
the energy when expressed in terms of its proper
independent variables, $E=E(S,A)$, is integrable.
We remark that all the above functions may be considered as functions either
of the cavity radius or the wall area, since obviously, $r = \sqrt{A/(4\pi)}$
and $dA = 8\pi r dr$.

\section{A Liquid Model for Black Holes.}

The starting point for attempting to model black hole thermodynamics in
terms of liquids is the statistical mechanical treatment of fluids. There,
it is known how to relate the various macroscopic 
thermodynamic properties of liquids (energy, pressure, temperature)
to the
internal, microscopic features such as the interaction/intermolecular
potential and the pair-correlation function. This latter function provides a
measure of the local structure of the fluid or gas. 
The typical approach to the
study of the liquid state starts with (a perhaps imperfect) knowledge of the
interatomic force law and the measured short range order (obtained
experimentally via X-ray or neutron scattering experiments). 
One then attempts to
infer the macroscopic or thermodynamic behavior of the liquid on the basis
of this microscopic information. By marked contrast, here we shall turn the
reasoning around and solve for the local ``structure'' and the ``interatomic''
potential of an analog (and possible fictitious!) 
fluid from explicit knowledge of black hole
thermodynamics. Two very important ingredients which will allow this
inverted procedure to be carried out in closed form are (i) the fact that
the thermal ensemble of black holes is spatially bounded and (ii) the fact
that we know the spatial dimension of this ensemble to be $d=2$.

The model fluid we shall deduce will be described in classical
terms. Let us use our knowledge of the spatial dimensionality of the black
hole ensemble at the outset.
To make a thermodynamic correspondence between black hole and fluid
means we seek a two-dimensional fluid whose partition function
over the $2N$-dimensional phase space is given
by (restoring the dependence on $\hbar$)
\begin{equation}
Z_N(\beta) = \frac{1}{(2\pi \hbar )^{2N} N !}
\int d\{p_i\} \, \int d\{r_i\} \, e^{-{\cal E}/kT}
\end{equation}
and where the total, nonrelativistic energy of a system of $N$ interacting point
particles of mass $m$ in $d=2$ is
\begin{equation}
{\cal E}(\{p_i\}, \{r_i\}) = \sum_{i \neq j}^{2N} \frac{p_i^2}{2m} +
U(\{r_i\}).
\end{equation}
Here, $U(\{r_i\})$ is the potential energy of the particle system 
which we assume to be
pairwise additive:
\begin{equation}
U = \frac{1}{2} \sum^{N}_{i \neq j} \phi(|{\bf r}_i - {\bf r}_j|).
\end{equation} 
This is always a reasonable assumption provided the fluid constituents have
no internal structure that couples to the potential.
{}From the theory of liquids, it is well known that an equation of state for the
isothermal compressibility $\kappa_T$ and the internal energy $E$
can be calculated
in terms of the
pair-potential $\phi$ and the pair-correlation function $g$ \cite{Goodstein}-
\cite{Ishihara},
to whit (restoring the dependence on $k_B$),

\begin{equation}
(a): \qquad
\rho \, k_BT \kappa_T = \rho \int_{system} d^2{\bf \tilde r}\, 
\left[ g(\tilde r) -1\right ] + 1, 
\end{equation}
and
\begin{equation}
(b): \qquad
 E = Nk_BT + \frac{1}{2}N \rho \, \int_{system} d^2{\bf \tilde r}\,\, 
\phi(\tilde r) g(\tilde r),
\end{equation}
where $\rho = \frac{N}{A}$ is the two-dimensional particle density, $T$ is
the fluid temperature, and 
the radial distribution function $g$
is defined via
\begin{equation}
\rho \, g(r) = \frac{1}{N}\langle \sum_{i \neq j} \delta ({\bf r} + {\bf r}_i 
- {\bf r}_j ) \rangle,
\end{equation}
the angular brackets denote the average computed using the grand
canonical ensemble.

Before we go on to use these relations, we remark that the
equation of state ($a$) is exact while the expression for the internal energy ($b$)
makes use of the pairwise summability of the total potential energy. 
They are valid for any single-component,
monatomic system in
thermodynamic equilibrium (gas, liquid or solid) 
whose energy is expressible in the form (28)
with a pairwise additive potential (29).  
Although these expressions 
are derived primarily for their application to the
liquid state, they can also be applied to the study of solids. The
only modification would be that $g$ and $\phi$ depend on the full vector
coordinate ${\bf r}$
(magnitude and direction). For liquids, however, the results are
isotropic so it is enough to write $g(r)$ and $\phi(r)$.  For the
present consideration, modelling a liquid which is capable of
reproducing certain aspects of black hole thermodynamics is carried
out once we identify the $\kappa_T$ and $E$ in ($a$) and ($b$) with those
of the black hole. 

The idea of representing a black hole at finite temperature
as a thermal fluid is novel and therefore deserves careful explanation. 
A black hole is but one example of a thermodynamic system having, among other
things, a well-defined temperature, energy and compressibility. On the other
hand, any equilibrium many-body system with hamiltonian given in Eq.(28) 
and Eq.(29) has a well-defined energy and compressibility which can be
calculated in terms of an associated $g$ and $\phi$. 
When we {\it formally} identify
the $\kappa_T$ and $E$ appearing there with those belonging to the black 
hole, we are simply demanding that these particular thermodynamic functions be
reproducible in terms of the internal variables of a certain classical
many-body system. This is not to say that these variables actually represent the true degrees of freedom of a black hole.
The identification must be carried out in a consistent way. 
First, the temperature $T$ appearing in (30) and (31) is the uniform cavity
wall temperature. Since ($a$) and ($b$) are to decribe a liquid, the temperature
of that liquid must be identified with this temperature: $T_{liquid} = T$.
Note that the  temperature of the liquid is {\em intensive}. That is, the 
temperature of the cavity wall of the black hole ensemble is 
identified with the temperature of the bulk fluid. 
Next, the density $\rho$ of the
fluid is simply the number of ``atoms'' per unit area of fluid. For the
black hole, both $E$ and $\kappa_T$ depend 
explicitly on the cavity radius $r$, reflecting the 
fact that the black hole ensemble is spatially finite. This means that the 
integrations in ($a$) and ($b$) are to be carried out over 
a fluid of bounded spatial extent.
The integrations over the $system$ are bounded.
Since the integrands are 
functions only of $r$ and $T$, we can therefore write
  
\begin{equation}
\int_{system} d^2{\bf  r} = \int_0^r { r}\,
d{ r}\int^{2\pi}_0 
d\theta.
\end{equation}

It is natural to take the length scale of the liquid coincident with that of
the cavity containing the black hole; any other choice would introduce 
a second, and arbitrary, length scale into the problem. 
Taking $E$ and $\kappa_T$ from (14) and (17) as input, the 
relations ($a$) and ($b$) 
yield two equations in the two unknowns
$g$ and $\phi$. 
We can easily solve for these microscopic functions in terms of
the macroscopic functions and their first derivatives. To do so, we
make use of (33) and differentiate both sides of the relations in 
($a$) and ($b$) with respect to $r$. The results of this operation are that
\begin{equation}
\phi(r,T) g(r,T) = \frac{4r}{N^2}\left[\left(\frac{\partial E}{\partial r}
\right)_T 
+ \frac{2}{r} \left( E - Nk_BT \right) \right],
\end{equation}
and 
\begin{equation}
g(r,T) - 1 = \frac{2r}{N} \left[ k_BT \left( 
\frac{\partial \rho \kappa_T}{\partial r} \right)_T + \frac{2}{r} \left( 
\rho k_B T \kappa_T - 1 \right) \right].
\end{equation}

By explicit construction, 
these give the pair correlation function and the inter-particle
potential of the model fluid whose energy and isothermal
compressibility are identical with those of the black hole. 
Moreover, these two functions depend on the two \underline{independent}
variables $r$ and $T$. Since we can vary them independently, we have actually
obtained $\phi$ and $g$ as functions of their arguments for all $T \geq 0$ and $r \geq 0$,
subject only to the restraint that the product $rT$ always be greater than 
or equal to $(rT)_{min}$. 
By way of a
trivial but illustrative example, consider the 
ideal gas in two-dimensions whose equation of
state is $pA = Nk_B T$. Then $E = Nk_BT$ and $\kappa_T
= -\frac{1}{A} (\frac{\partial p}{\partial A})_T = 
1/(\rho k_BT)$. 
Inserting these into the above relations 
immediately yields $g(r) = 1$ and 
$\phi(r) = 0$, which is also a solution of the pair of 
equations (30) and (31). 
As is to be expected, the ideal gas has no 
structure (it is uniform: homogeneous and isotropic) and lacks 
interatomic interactions (by definition). Therefore, any deviation in either
$g$ and or $\phi$ with respect to these limits may be considered as deviations
from an ideal gas.

It is of interest to consider the limiting forms of the pair correlation
function and potential energy for the black hole; these may be deduced 
easily from the associated
limits calculated above for $E$ and $\kappa_T$, Eqs.(18,19) and Eqs.(22,23).
When $rT >> (rT)_{min}$, the pair correlation function goes as
\begin{equation}
(i) \qquad g(r,T) \sim 1 - \frac{8k_B T}{r} - \frac{4}{N}.
\end{equation}
In particular, for fixed temperature, $g(r,T) \rightarrow 1 - \delta$, as
$r \rightarrow \infty$
where $\delta = 4/N$ is small for $N$ large. 
In normal simple liquids, $g(r)$ has
the asymptotic limit $g(r) \rightarrow 1$ (compare to the ideal gas limit) and
deviations from this value represent molecular correlations 
(or anti-correlations). When evaluated along the boundary
hyperbola $rT = (rT)_{min}$ we get, 
\begin{equation}
(ii) \qquad g(r,T) = 1 - \frac{4}{N},
\end{equation}
a constant independent of $r$ and $T$.  The corresponding 
limits for the two-body potential energy may be worked out and yield
\begin{equation}
(i) \qquad \phi(r,T) \sim \frac{4r}{N^2}\left(3 - \frac{2Nk_BT}{r} \right)/
\left(1 - \frac{8k_B T}{r} - \frac{4}{N} \right).
\end{equation}
For fixed temperature, $\phi \sim r$. When $rT = (rT)_{min}$, 
\begin{equation}
(ii) \qquad \phi(r,T) = \frac{4r}{N^2} \left[(3 - \sqrt{3}) - \frac{2Nk_BT}{r}
\right]/\left(1 - \frac{4}{N}\right).
\end{equation}

The black hole pair correlation function is calculated and presented in Fig. 4.
For fixed $T$ and small $r$, this function is negative, then increases, 
becoming positive and approaches unity from below as $r \rightarrow \infty$. 
This behavior is also revealed in the one dimensional plot of $g(r,T)$ for
the value $T = 0.5$ in Fig. 5. What can we make of this 
behavior in $g$ and what physical interpretation can it admit? For this, let us
turn to the meaning of $g(r)$. Imagine we select a particular particle of the
fluid, whose average density is $\rho$, 
and fix our origin at that point. Then, the number of particles $dN$
contained within the (two-dimensional) spherical shell of thickness
$dr$ centered at $r=0$ is
\begin{equation}
dN = \rho\, g(r) 2\pi r\, dr.
\end{equation}
Here we see that $g$ gives a measure of the deviation from perfect homogeneity
($g = 1$). Evidently, $ g < 0 \iff dN < 0$ in that shell. On the other hand, a
negative value for $dN$ is the signature for the phenomenon of 
charge-screening, i.e., it indicates the presence of {\it holes} in the
neighborhood of our reference particle at $r=0$. Thus, it would appear that 
the analogue fluid which could model some aspects of black hole thermodynamics
should have something to do with a charged fluid or plasma. 
These latter systems are
defined as a collection of identical point charges, of equal 
mass $m$ and charge $e$, embedded in a uniform background of charge (the dielectric) obeying
classical statistical mechanics. If one adds a given charge to the plasma, the
plasma density is locally depleted so as to neutralize the impurity charge. 
This is the well-known phenomenon of charge screening. The depletion shows up
as an underdensity of particles (or an overdensity of holes), and is reflected
in a $g<0$ near the origin, that is, where the impurity charge is located. 
Calculations of the pair-correlation function
for degenerate electron plasmas at metallic densities yield functions exhibiting the same general qualititative features as those in Figures 4 and 5
\cite{March}. In addition, screening is known to be a characteristic property of interactions like the electromagnetic interaction, where there exist two species of charges; they have the property that renormalization effects induce corrections that make the $effective$ charge decrease with distance. This, in itself, is not surprising because as is well known there exists an analog of a black--hole with an electromagnetic membrane; an analogy that in the literature is called the ``membrane paradigm" ~\cite{thorneprice}. What we find here is a different manifestation of this analogy, this time through the dynamical and statistical properties of the liquid.

The weakly temperature dependent potential is scaled by $N^2$ and plotted
in Figure 6. Again, recall the physical part of the graph consists of those
points $r$ and $T$ satisfying $rT > (RT)_{min}$. The potential is seen to
be a positive increasing function of $r$. The limit calculated above in
Eq.(38) shows the growth is essentially linear. Apart from the ``glitch''
near $T \approx 0.2$ the potential is practically independent of the
temperature.

\section{Conclusions.}

It is worth emphasizing that the analogue fluid selected to account
for the black hole compressibility and internal energy was ``engineered''
at the fluid's atomic level. As there is no corresponding ``atomic''
level for the black hole, the bridge between the thermodynamics of the
black hole and liquid is established via thermodynamic state functions.
Surprisingly, only two state functions are needed in order to specify
completely the ``atomic potential" and local structure of the analogue
fluid. However, as we have seen, we can only be sure that the fluid will reproduce
the correct compressibility and internal energy. That is, only
partial aspects of black hole thermodynamics will be reproducible, since,
evidently, there exist other state functions that characterize a 
black hole, namely, its entropy, pressure, specific heats, etc. which must be calculable by a more complete ``microscopic" description of the black hole (see below). The analogy with the liquid leads to a screening effect that can be understood in terms of a connection with the membrane paradigm.

The partial rendering of black hole thermodynamics in terms of atomic
fluid elements achieved here points to the possibility of directly effecting
a mapping between the black hole variables (mass $M$ and cavity radius
$r$, or cavity wall temperature $T(r)$) and the internal variables
of an analogue model which might serve to reproduce all of 
black hole thermodynamics. Evidently, this would amount to a formal
correspondence at the level of the degrees-of-freedom and so bypass
the need to call into play the macroscopic state functions. A concrete
example of such a mapping between two entirely distinct systems
is that established recently between 
a Newtonian cosmology of pointlike galaxies and spins in a
three-dimensional Ising model \cite{PM-etal}. The degree-of-freedom
mapping problem is well
worth pursuing as intriguing deep connections between gravitation,
thermodynamics and information theory have been hinted at recently
\cite{BrownYork}. Another hint is supplied by Wheeler's depiction of
the Bekenstein bit number as a ``covering'' of the event horizon by
a binary string code representing the information contained in the
black hole \cite{Wheeler1,Wheeler2}. It may well be possible to go beyond
these provocative hints and actually establish a rigorous connection
between black holes, computation, information theory and complexity.
We hope to report on these developments in a separate paper.

\newpage
\begin{center}
Figure Captions.
\end{center}

Figure 1. The black hole energy, $E$, as a function of the cavity radius $r$ and cavity wall temperature $T$. It is seen that $E$ grows with cavity radius and with temperature. The flat portion of the graph (in this and in the following figures) is not part of the physical quantity shown in the graph; it corresponds to the region  $rT<(rT)_{min}$ in which the black hole mass is no longer real and positive.
\vspace{.5cm}

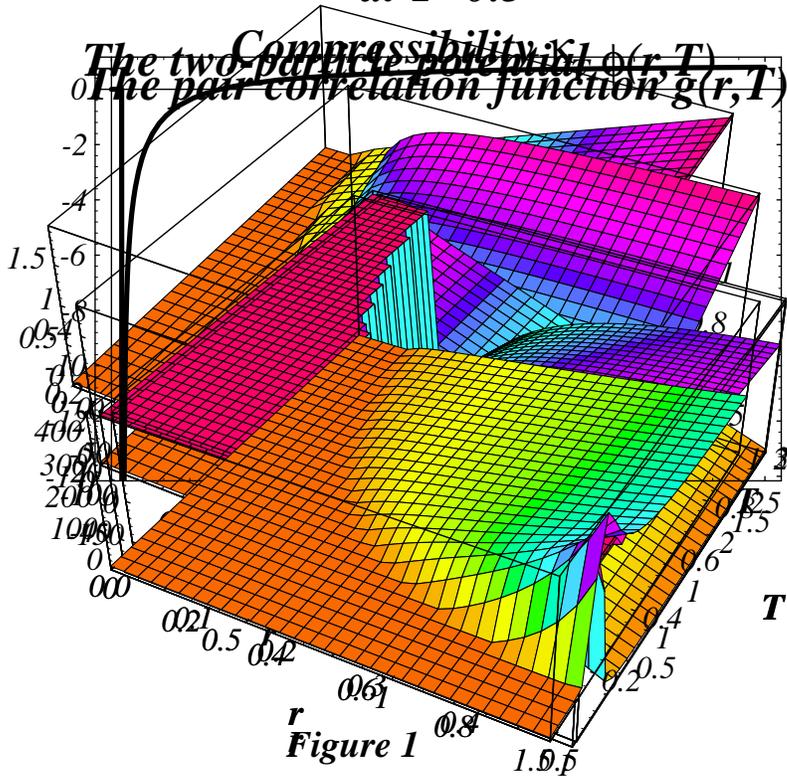
Figure 2. The black hole surface pressure, $\sigma$, as a function of the cavity radius $r$ and cavity wall temperature $T$. It is seen that $\sigma$ grows sharply at first with cavity radius to then stabilize its growth, while its growth with temperature is unimpeded. The flat portion of the graph (as in all figures in this paper figures) is not part of the physical quantity shown in the graph; it corresponds to the region  $rT<(rT)_{min}$ in which the black hole mass is no longer real and positive.
\vspace{.5cm}

Figure 3. The black hole isothermal compressibility, $\kappa_T$, as a function of the cavity radius $r$ and cavity wall temperature $T$. It is seen that $\kappa_T$ decreases with cavity radius and with temperature. The flat portion of the graph (as in all figures in this paper figures) is not part of the physical quantity shown in the graph; it corresponds to the region  $rT<(rT)_{min}$ in which the black hole mass is no longer real and positive.
\vspace{.5cm}

Figure 4. The black hole pair correlation function, $g$, inferred in the liquid model of the black hole as a function of the cavity radius $r$ and cavity wall temperature $T$. It is seen that for $r>.2$, $g$ grows with cavity radius but decreases with temperature. The flat portion of the graph (in this and in the following figures) is not part of the physical quantity shown in the graph; it corresponds to the region  $rT<(rT)_{min}$ in which the black hole mass is no longer real and positive. The spikes are {\em numerical artifacts} of the plotting routine used to plot the figure and which could not render the deep trough from which the physical region emerges.
\vspace{.5cm}

Figure 5. Two dimensional representation of the black hole pair correlation function, $g$, as a function of the cavity radius $r$ but for a fixed value of the cavity wall temperature $T=0.5$. This is plotted here to illustrate the existence of the trough which was not clearly seen in the three dimensional plot of Figure 4. Notice that for large $r$ the pair correlation function tends to the constant value +1. 
\vspace{.5cm}

Figure 6. The  black hole ``liquid" two--particle potential, $\phi$, as a function of the cavity radius $r$ and cavity wall temperature $T$. It is seen that apart from the sudden rise near $T=0.2$, the potential is a rather boring function of cavity radius and temperature. The flat portion of the graph (in this and in the previous figures) is not part of the physical quantity shown in the graph; it corresponds to the region  $rT<(rT)_{min}$ in which the black hole mass is no longer real and positive.
\end{document}